# Article Title: Bayesian estimation methods for survey data with potential applications to health disparities research

**Article Category:** Advanced Review

**Authors:**

| |
|---|
| **Stephanie M. Wu** <br> Harvard T.H. Chan School of Public Health, swu@g.harvard.edu, https://orcid.org/0000-0001-5110-8407 |
| **Briana Joy K. Stephenson\*** <br> Harvard T.H. Chan School of Public Health, bstephenson@hsph.harvard.edu, https://orcid.org/0000-0002-6147-1039 |

**Abstract**

Understanding how and why certain communities bear a disproportionate burden of disease is challenging due to the scarcity of data on these communities. Surveys provide a useful avenue for accessing hard-to-reach populations, as many surveys specifically oversample understudied and vulnerable populations. When survey data is used for analysis, it is important to account for the complex survey design that gave rise to the data, in order to avoid biased conclusions. The field of Bayesian survey statistics aims to account for such survey design while leveraging the advantages of Bayesian models, which can flexibly handle sparsity through borrowing of information and provide a coherent inferential framework to easily obtain variances for complex models and data types. For these reasons, Bayesian survey methods seem uniquely well-poised for health disparities research, where heterogeneity and sparsity are frequent considerations. This review discusses three main approaches found in the Bayesian survey methodology literature: 1) multilevel regression and post-stratification, 2) weighted pseudolikelihood-based methods, and 3) synthetic population generation. We discuss advantages and disadvantages of each approach, examine recent applications and extensions, and consider how these approaches may be leveraged to improve research in population health equity.

Keywords: Bayesian statistics, health disparities, survey design, population health

**Graphical/Visual Abstract and Caption**

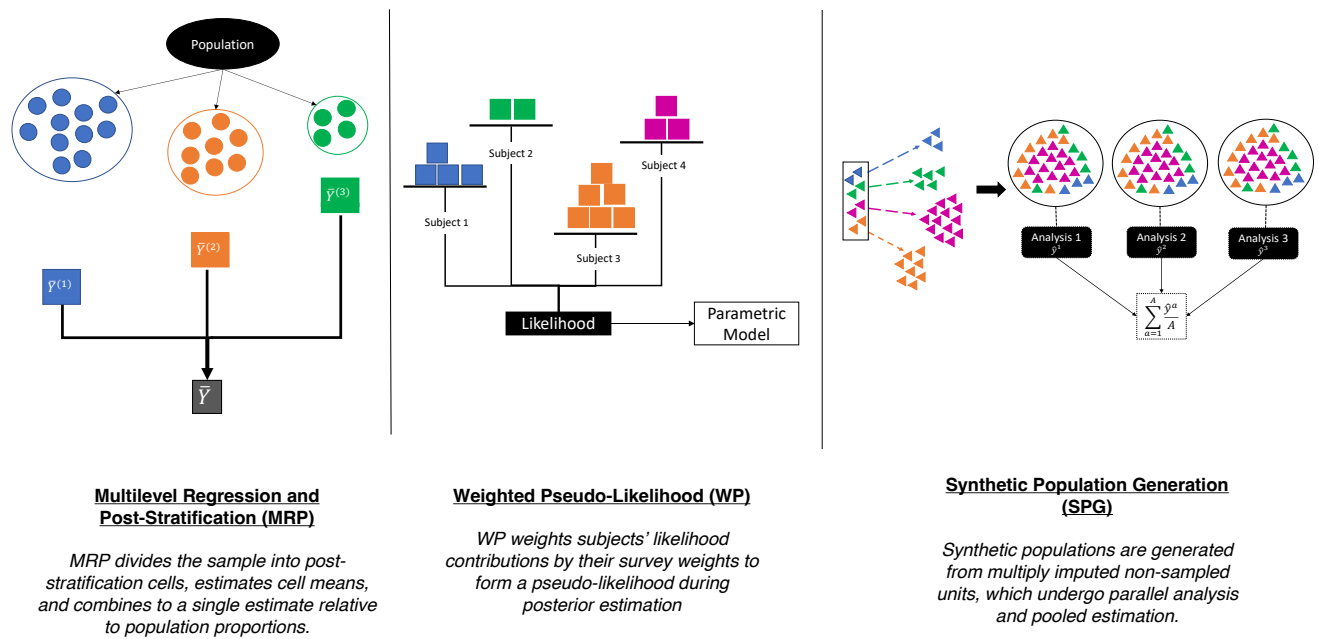

*Figure 1: Visual summary of three Bayesian survey methodology approaches explored in this review, with potential applications in health disparities research.*

# 1. INTRODUCTION

Health disparities, as defined by the Centers for Disease Control and Prevention (CDC), are "preventable differences in the burden of disease, injury, violence or opportunities to achieve optimal health by socially disadvantaged populations" (Centers for Disease Control and Prevention, 2008). Socially disadvantaged populations are identified by an unequal distribution of representation and resources. Social disadvantages prevalent in the United States include considerations of race, ethnicity, gender, economic/financial status, geographic location, and physical disabilities.

In national studies, these same populations are often underrepresented, and are consequently understudied due to statistical limitations in sample size and power (Shavers-Hornaday et al., 1997). This is perhaps most clear in clinical research, where patient populations such as racial and ethnic minority groups, females, and the elderly continue to be under-enrolled despite often carrying a disproportionate burden of health conditions of interest (Ortega et al., 2019; Tahhan et al., 2020). Survey sampling methodologies have attempted to overcome these limitations by changing sampling strategies to over- or under-sample certain demographics so there is enough data capturing the heterogeneity of the population. For example, the National Health and Nutrition Examination Survey has evolved their sampling strategies over the past several decades to adapt to the constantly growing cultural and demographic diversity of the United States, oversampling persons aged 60 and older, living at or below poverty level, and/or identifying as African-American, Hispanic/Latinx-American, or Asian-American (Johnson et al., 2014). Such public use datasets are typically accompanied with survey weights, which account for unequal selection probabilities and other sources of selection bias. Surveys focused on rare populations that are stigmatized and/or hard to reach due to demographic scarcity may use non-probability sampling strategies such as respondent-driven sampling (Heckathorn, 1997). For example, the National HIV Behavioral Surveillance System implements this strategy to sample through three rotating demographic cycles: men who have sex with men, persons who inject drugs, and heterosexually active persons at increased risk for HIV (Gallagher et al., 2007). In these surveys, the selection mechanism and weights are unknown and must be estimated.

Surveys play an important role in gathering data on underrepresented and stigmatized groups and accurately capturing the diversity present in larger populations – diversity that is necessary for understanding health disparities and population health differences. To produce reliable estimates and generate appropriate population inference, analyses must acknowledge the survey design to allow the sample to reflect the target population more closely. It is paramount for population health analysis to have reliable statistical methods that can properly examine these heterogenous groups and provide generalizable results to the broader population.

Advances in computation and data accessibility have created a recent boom in advanced statistical methods that can accommodate big data and address population health challenges. However, the sampling strategy used in acquiring the available data is not always responsibly considered during methodology development. Statistical methods for survey data have historically been developed through a frequentist paradigm of design-based inference. A Bayesian model-based framework can offer advantages in survey statistics through flexibly adapting sparse and complex data and providing a coherent inferential framework, but Bayesian methods often ignore survey design and assume observations are independent and identically distributed. The integration of these two fields has great implications for improving our ability to accurately use survey data to understand understudied populations.

Within the sphere of health disparities research, implementation of Bayesian methods with survey data has been primarily focused on small area estimation to obtain summary estimates of different subgroups in a population, with vast applications in global public health and spatial statistics (Burstein et al., 2019; Dwyer-Lindgren et al., 2016; Gething et al., 2016; Gutreuter et al., 2019; Mercer et al., 2015; Utazi et al., 2021). Several statistical reviews and texts have covered these methods extensively for small area estimation and model-based geostatistics (Battese et al., 1988; Diggle & Giorgi, 2019; Fay & Herriot, 1979; Ghosh, 2020; Parker et al., 2023; Rao & Molina, 2015; Wakefield et al., 2020). While more innovative Bayesian model-based approaches have been recently developed and applied in population health research to address subgroup heterogeneity and data sparsity, few have been translated to accommodate survey research settings outside of the spatial statistics domain (Alexander et al., 2017; Stephenson & Willett, 2023; Yang & Puggioni, 2021).

This review aims to provide a generalized overview of current Bayesian survey approaches and their applications, and to discuss the potential of integrating these methods with recent Bayesian model-based techniques to improve population health disparities research. We organize this review as follows: Section 2 introduces major concepts and notation. Section 3 discusses the use of multilevel regression and post-stratification. Section 4 discusses approaches that rely on a weighted pseudolikelihood. Section 5 discusses approaches that generate synthetic populations. Section 6 examines how we can use the current approaches to improve health disparities research and future directions. We conclude with a discussion and summary in Section 7.

## 2. MAJOR CONCEPTS AND NOTATION

### 2.1 Analysis of survey data

Standard statistical methods assume independent and identically distributed (IID) data, which corresponds to simple random sampling with replacement in a finite population sampling framework. Survey data, however, typically arise from complex sampling designs that include features that influence selection into the sample, such as unequal probabilities of selection, cluster sampling, and stratification of the population. These features are implemented for practical convenience and to ensure representation of key subgroups within a population. Failure to account for complex sampling design features can result in substantially biased estimation and inference (Heeringa et al., 2017; Skinner, 1989; West et al., 2016). Survey weights, often provided with survey data, encode these complex design features to adjust for selection bias and can also include adjustments for nonresponse bias and calibration to known population totals (Lohr, 2021), and are therefore useful to incorporate for valid inferences using survey data.

### 2.2 Design-based vs. model-based inference

Traditionally, survey methods have been developed under the frequentist design-based framework and have focused on descriptive inference for population means and totals. There are two main approaches to survey data inference: design-based inference and model-based inference. Design-based inference assumes the finite population data is fixed and randomness arises from the sample inclusion mechanism, which induces the statistical distribution of an estimator. Here, models are not explicitly assumed, though they may implicitly assist estimators that are chosen based on desirable asymptotic properties (e.g., design-consistency and efficiency). Issues with design-based inference can arise with uncertainty estimation, small sample sizes, and adjustments for nonresponse errors. In these cases, model-based inference can be advantageous (Little, 2004).

Model-based inference assumes the population is not fixed and that it is a realization of a random data generating process that is described via a parametrized model. Here, estimation and inference are based on the model rather than on the probability sampling distribution, but probability sampling is still useful for accounting for selection bias (Gelman et al., 2013; Little, 2004). Model-based inference can be frequentist, through considering repeated samples from a "superpopulation" model (Royall, 1970), or Bayesian, by placing a prior distribution on the population parameters and performing inference based on the posterior distribution (Ericson, 1969). Bayesian model-based inference has become popular in recent years and includes important advantages for small-sample inference. It also offers a unified approach for handling sampling design, non-response, and combining data sources. However, valid inference depends on correct specification of the model, which is difficult to verify, and modelling all survey variables can be involved or impossible if the relevant information is not publicly available (Little, 2004).

Employing a hybrid "calibrated Bayes" framework offers a means to harmonize design- and model-based approaches for sample survey inference (Box, 1980; Little, 2006, 2012, 2014; Rubin, 1984). Under this framework, inference is based on a Bayesian model, but the model is chosen to yield inferences that are well-calibrated in a frequentist sense, such as producing unbiased point estimates and posterior probability intervals with (approximately) nominal frequentist coverage. This paradigm leverages the optimality of Bayes under a correctly specified model, while prioritizing inferences that are robust to misspecification and have good repeated-sampling characteristics. The methods described in this review are considered and compared under the calibrated Bayes framework. For a review of available software under the various survey frameworks, see West et al. (2018).

## 2.3 Notation

For the remaining sections of this review, we use the following notation to describe the survey data. Let $y_1, ..., y_n$ denote the observed outcome of interest in a sample of size $n$ from a finite population of size $N$, and let $w_1, ..., w_n$ denote the corresponding available survey weights, assumed to be inversely proportional to the marginal inclusion probabilities and normalized to sum to $N$. If a parametric model is assumed, its density is denoted by $p(y|\theta)$, with $\theta$ being the parameter(s) of interest. Information on the selection mechanism and survey design may be limited to the researcher at the time of analysis. We consider the scenario where the only information available to the analyst is the final survey weights, and we specify when additional data such as cluster and stratum indicators are necessary or helpful for improved estimation and inference.

## 3. MULTILEVEL REGRESSION AND POST-STRATIFICATION FOR SUBGROUP ESTIMATION OF MEANS

## 3.1 Background and overview

Multilevel regression and post-stratification (MRP), first proposed by Gelman & Little (1997), is commonly used to estimate means and totals of population subgroups while accounting for selection bias. MRP has two main components: 1) a Bayesian multilevel regression model for small area estimation outcome prediction, and 2) post-stratification to adjust for survey design. In small area estimation (SAE) (Rao & Molina, 2015), the goal is to obtain reliable estimates, usually of the mean, of "small areas" – domains such as geographic areas or demographic subgroups that have few or no samples available and therefore cause issues with sparsity and stability of parameter estimates. Bayesian multilevel/hierarchical regression models are a popular tool among SAE methods for allowing information to be borrowed across subgroups and stabilizing estimates.

Post-stratification (Holt & Smith, 1979) corrects for differences between sample and population that can arise from survey design elements such as under- and over-sampling and nonresponse. Post-stratification partitions the population into cells, formed by cross-tabulation of all variables that are predictive of the outcome and/or influence survey design (e.g., county, race, education level), converted into categorical form. Within each of these post-stratification cells, the survey can be treated as simple random sampling since all design variables are used when creating the cells. Post-stratification uses auxiliary population information on these design variables to re-weight observations in the sample to match population values, thereby removing the effect of survey design.

MRP uses a model-based perspective of post-stratification (Little, 1993a). First, the outcome mean within each post-stratification cell is estimated using a regression model of the outcome conditional on the design variables. This falls under a superpopulation modeling framework, where information about non-sampled units is predicted by assuming some underlying model holds for generating both the sampled and non-sampled data, conditional on variables that influence survey design. A multilevel regression model is typically used here to provide more stable small area estimates. Then, the population mean is estimated using a weighted average of all the post-stratification cell mean estimates, where each cell's contribution is proportional to its (known or estimated) size in the population so that the sample is re-weighted to match the population composition.

## 3.2 Description of multilevel regression and post-stratification approach

Classical MRP corrects for population-sample differences by leveraging additional data sources, such as administrative data or census records, that have population-level data on auxiliary variables, such as age and sex, that influence selection and response; these auxiliary variables are then used to form post-stratification cells (Gelman and Little, 1997; Gao et al., 2021). Si et al. (2015) propose a robust MRP approach for estimation of means and proportions using survey data. Since variables influencing design are typically integrated into the survey weights, Si et al. (2015) remove the need to use auxiliary population data and instead use the unique values of the observed survey weights to define the post-stratification cells, resulting in all those in the same cell sharing the same weight. It is assumed that all cells in the population are represented in the sample.

The approach relies on the following setup: for each unit in the sample, we observe an outcome $y_i$ and survey weight $w_i$, $i = 1, \ldots, n$, where $n$ is the sample size. Suppose the survey weights take on a total of $J$ different values, resulting in $J$ post-stratification cells. For each cell $j = 1, \ldots, J$, let $n_j$ denote the sample cell size and $N_j$ denote the known or unknown population cell size, with $n = \sum_{j=1}^{J} n_j$ and $N = \sum_{j=1}^{J} N_j$. The parameter of interest is the population mean, $\theta = \bar{Y}$, and the MRP estimator for $\theta$ is constructed as the weighted average of the post-stratification cell means:

$$\hat{\theta} = \frac{\sum_{j=1}^{J} \hat{N}_j \hat{\theta}_j}{\sum_{j=1}^{J} \hat{N}_j} = \frac{\sum_{j=1}^{J} \left( n_j \bar{y}_j + (\hat{N}_j - n_j) \bar{y}_{\text{non},j} \right)}{\sum_{j=1}^{J} \hat{N}_j}. \tag{3.1}$$

Here, $\hat{\theta}_j$ is the estimate for the true mean of cell $j$, $\theta_j$. Each cell mean $\hat{\theta}_j$ is estimated using a multilevel model conditional on the survey weights. This model is used to relate $\bar{y}_j$, the cell mean from the sampled units, and $\bar{y}_{\text{non},j}$, the cell mean estimate for the non-sampled units, which must be predicted. $\hat{N}_j$ is the estimate for the population cell size $N_j$ and is necessary for re-weighting the post-stratification means. If $N_j$ is known (e.g., from some external data source), then it can be plugged into equation (3.1) directly.

*Multilevel regression model:* To form the outcome regression model and obtain $\hat{\theta}_j$ and $\bar{y}_{\text{non},j}$, Si et al. (2015) propose using a Bayesian non-parametric multilevel regression model for the outcome conditional on the weights and other predictors of the outcome. A Gaussian process (GP) prior is placed on the conditional mean $\mu(\cdot)$ to minimize possible bias from the regression model. Let $w_j$ denote the shared survey weight for cell $j$. For a continuous outcome $y_i$, for all units $i$ in cell $j$, we have

$$\begin{aligned} y_i &\sim N\left(\mu(\log(w_j)), \sigma^2\right) \\ \mu(\log(w_j)) &\sim GP\left(\log(w_j)\beta, \Sigma\right), \end{aligned} \tag{3.2}$$

where $\sigma^2$ is the individual variance, $\beta$ denotes an unknown coefficient, and $\Sigma$ denotes the covariance function that controls the amount of borrowing of information from neighboring cells. Note that different outcome types will yield different models based on the appropriate distribution; for example, binary outcomes would use a binomial model. Spatial random effects can also be included for more granular small area estimation (Vandendijck et al., 2016). Alternative flexible models, such as penalized splines (Q. Chen et al., 2010; Zangeneh & Little, 2015; Zheng & Little, 2005, 2003), complex hierarchical models with interactions (Ghitza & Gelman, 2013; Goplerud, 2023), and ensemble machine learning models (Bisbee, 2019; Broniecki et al., 2022; Ornstein, 2020) can also be used to form the outcome regression model but are not explored in this review.

*Post-stratification:* To adjust for survey design, the population cell size $N_j$ is necessary and must be estimated if it is not known. $\widehat{N}_j$ can be estimated as a parameter corresponding to multinomial likelihood

$$(n_1, \ldots, n_J) \sim \text{Mult}\left(n; \frac{N_1/w_1}{\sum_{j=1}^J N_j/w_j}, \ldots, \frac{N_J/w_J}{\sum_{j=1}^J N_j/w_j}\right), \tag{3.3}$$

where $n$ is the fixed sample size. Note that estimating $\widehat{N}_j$ implicitly estimates the weights for the non-sampled units and the re-weighting that occurs through post-stratification (Si et al., 2015).

*Parameter estimation:* After placing appropriate priors on all parameters, posterior estimates can be obtained for $\widehat{N}_j$ and the regression model parameters. Then, $\bar{y}_{\text{non},j}$ can be predicted by using the mean of the posterior predictive distribution. The MRP estimate for the population mean follows by using equation (3.1). Estimates for subgroup means can be obtained over aggregates of a subset of the post-stratification variables. Posterior inference is generally robust under different prior distributions, with weakly informative priors typically used as a default in the absence of prior information (Si et al., 2015). Prior specification can also be used to control the amount of pooling among post-stratification cells, as well as improve precision by taking advantage of informative structure in the data (Gao et al., 2021; Si et al., 2020).

## 3.3 Advantages and disadvantages

Advantages of MRP include accurate and efficient performance for small area estimation, allowing for partial pooling across sparse cells and outperforming classical design-based estimators in terms of stability, efficiency, and coverage in comprehensive simulation studies (Si et al., 2015). It is also relatively straightforward to implement using the Stan programming language (Carpenter et al., 2017), is computationally efficient, and does not require additional design information beyond survey weights, though cluster membership indicators are necessary when cluster sampling adjustments are to be applied (Makela et al., 2018).

Disadvantages include possible bias if the sample inclusion mechanism depends on non-sampled outcomes, equivalent to violating the ignorable missingness assumption (Rubin, 1987). If not all post-stratification cells in the population are represented in the sample, MRP requires additional pooling to predict outcomes and weights of empty cells, with one approach being to model all the survey weights and use structured prior distributions to smooth and stabilize estimation (Si et al., 2020). In addition, model specification can be challenging in scenarios with complicated data types or many post-stratification variables (Si et al., 2015). Consequently, current inference is mostly limited to simple descriptive inference of population and subgroup means and totals.

### 3.4 Applications and extensions

Applications and extensions of MRP, summarized in **Table 3.1**, span multiple substantive domains, including health and well-being (Covello et al., 2021; Si et al., 2020; Vandendijck et al., 2016; Watjou et al., 2017), poverty (Makela et al., 2018; Si et al., 2015; Si & Zhou, 2021), employment (Downes & Carlin, 2020), education (Zangeneh & Little, 2022), and public opinion and election forecasting (Gao et al., 2021; Ghitza & Gelman, 2020; Lauderdale et al., 2020; Leemann & Wasserfallen, 2017). Methodologically, the MRP approach is used for descriptive inference of means, proportions, and totals at the population and subgroup level. Because MRP can account for survey design while maintaining stable estimates across sparse geographic and/or demographic subgroups, it allows estimates to be obtained for understudied communities that may have limited data available due to historical exclusion, lack of accessibility, new migration trends, or other reasons, and for which the large sample size requirements of traditional survey analysis methods are not met. However, these analyses are restricted to simple summary estimates, and methods are still lacking for adapting MRP to more complex Bayesian models.

**Table 3.1 Summary of applications and extensions using MRP approach**

| | Author(s) | Data application | Survey type | Statistical method |
|---|---|---|---|---|
| Health and well-being | Covello et al. (2021)** | Prevalence of asymptomatic SARS-CoV-2 in Indiana urban-suburban-rural setting | Routine hospital-based testing of patients for elective medical procedures | Population proportion estimation over time with calibration to target population demographics |
| | Si et al. (2020)** | Average score of life satisfaction overall and for subgroups of adults in NYC | Longitudinal Survey of Wellbeing (LSW): unequal probabilities of selection, nonresponse, calibration | SAE of continuous outcome. Extension to weight generation using hierarchical priors for robustness and stability |
| | Vandendijck et al. (2016)** | Prevalence of asthma across districts in Belgium | Belgian Health Interview Survey: stratified, multi-stage cluster sampling | Extension to SAE of proportions. Use random effects to account for spatial distribution |
| | Watjou et al. (2017)* | Prevalence of good perceived health across districts in Belgium | Belgian Health Interview Survey: stratified, multi-stage cluster sampling | SAE of proportions. Extension to spatial smoothing methods and impact of nonresponse |
| Poverty | Makela et al. (2018)* | N/A (synthetic outcome) | Fragile Families and Child Wellbeing: modified to two-stage cluster sampling with highly variable weights | Population mean estimation for continuous or binary outcome. Account for cluster sampling by predicting unknown cluster sizes |
| | Si et al. (2015) | Mean public assistance, welfare, and food stamp use among mothers of nonmarital births in NYC | Fragile Families and Child Wellbeing Study: unequal selection, nonresponse, calibration | Population mean estimation for continuous or binary outcome |
| | Si & Zhou (2021)** | Prevalence of material hardship overall and for subgroups of adults in New York City (NYC) | Longitudinal Survey of Wellbeing (LSW): unequal probabilities of selection | SAE of proportions. Extension to Bayes-raking. Calibration with marginal constraints through priors on population cell counts |
| Employment | Downes & Carlin (2020)* | Mean labor force status and hours worked at the national and state level | Ten to Men: The Australian Longitudinal Study on Male Health: stratified, multistage cluster sampling | Population mean estimation for continuous or binary outcome |
| Education | Zanganeh & Little (2022)* | Academic performance index score of schools in California with missing data | Academic Performance Index: simple random sample | Estimation of population mean using external post-stratification information, allowing for missing not at random unit missingness |
| Public opinion and election forecasting | Gao et al. (2021)** | Proportion of favorable opinions towards same-sex marriage among adults in US | National Annenberg Election Survey Phone Edition, American Community Survey | Small area estimation (SAE) of proportions. Bias and variance reduction using structured priors |
| | Ghitza & Gelman (2020)* | Proportion of voters supportive of incumbent, overall and by county, for 2012 US presidential election | Catalist voter registration database, Greenberg Quinlan Rosner Research poll, Democracy Corps poll | SAE of proportions. Incorporation of large-scale voter registration databases |
| | Lauderdale et al. (2020)* | US and UK election forecasting for sub-national electoral units | Pre-election polling, past election results, census data | SAE for categorical outcome |
| | Leemann and Wasserfallen (2017)** | Subnational proportions of voters in favor of Swiss People's Party referendums and initiatives | Switzerland VOX surveys with limited census data | SAE of proportions. Extension to MRP with synthetic joint distributions when only marginal variable distributions are known |

*: software referenced. **: software referenced and code available

# 4. WEIGHTED PSEUDO-LIKELIHOOD FOR COMPLEX PARAMETRIC MODELS

## 4.1 Background and overview

Weighted pseudo-likelihood (WP) approaches, often applied to complex parametric models, account for survey design through weighting the likelihood using available survey weights, then using the weighted pseudo-likelihood to construct a pseudo-posterior distribution for parameter estimation. WP methods evolved from frequentist pseudo-likelihood analyses (Binder, 1983; Skinner, 1989), where a pseudo-log-likelihood is formed by multiplying each individual log-likelihood unit in the sample by the respective survey weight:

$$\sum_{i=1}^{n} w_i \log(p(y_i|\theta)). \quad (4.1)$$

Estimation is obtained by maximizing the pseudo-log-likelihood in (5.1). Variance estimation and inference is often obtained through resampling methods such as balanced repeated replication (Kolenikov, 2010). The pseudo-likelihood is also referred to as a *composite likelihood* in general statistical problems where the weights are assumed to be fixed and known positive constants (Varin et al., 2011). Because the pseudo-likelihood does not incorporate the dependence structure in the sampled data or the variability in the survey weights, it is not a genuine likelihood. However, it is a useful tool for likelihood analysis for finite population inference. For WP approaches under a Bayesian framework, the pseudo-likelihood can be paired with a proper prior distribution and give rise to parameter estimation through the resulting weighted pseudo-posterior.

For the pseudo-log-likelihood in (5.1), the corresponding pseudo-score equation is a type of survey-weighted estimating equation, whose solution provides an estimator for a superpopulation model parameter $\theta$, such as the population mean, that is the solution to the corresponding "census" estimating equations formed using data from all population units (Binder & Patak, 1994). Many estimators can be formulated in this way, including the Horvitz-Thompson estimator, the ratio estimator, and regression coefficients (Binder & Patak, 1994). Though not highlighted in this review, there are several promising Bayesian pseudo-empirical-likelihood methods within the survey-weighted estimating equation framework that have been shown to have desirable asymptotic properties for estimation of parameters including the population mean (Rao & Wu, 2010), quantile regression coefficients (Zhao, Ghosh, et al., 2020), and the Gini coefficient for income inequality (Zhao, Haziza, et al., 2020). Survey-weighted estimating equations also feature in semiparametric models, such as Yiu et al. (2020)'s Bayesian exponentially tilted empirical likelihood method and Wang et al. (2018)'s approximate Bayesian generalized method of moments approach, both of which exhibit nice asymptotic properties.

## 4.2 Description of weighted pseudo-likelihood approach

Savitsky & Toth (2016) present the survey-weighted pseudo-posterior approach, incorporating the pseudo-likelihood into a Bayesian framework. The finite population data $(y_1, \ldots, y_n)$ are assumed to be generated independently from some parameterized model $p(y|\theta)$, where $\theta$ denotes the population parameter(s) of interest. Assume survey weights $(w_1, \ldots, w_n)$ that are inversely proportional to the marginal inclusion probabilities are available. These survey weights are treated as fixed, making this a plug-in approach.

The pseudo-posterior approach involves normalizing the weights to sum to the sample size to account for posterior estimation uncertainty, where $(\widetilde{w}_1, \ldots \widetilde{w}_n)$ denotes the set of normalized weights, $\sum_{i=1}^{n} \widetilde{w}_i = n$. This weighted pseudo-likelihood replaces the sample likelihood to construct a sampling-weighted pseudo-posterior, given by

$$p(\theta|y_1, \ldots, y_n, \widetilde{w}_1, \ldots \widetilde{w}_n) \propto p(\theta) \prod_{i=1}^{n} \{p(y_i|\theta)^{\widetilde{w}_i}\}. \quad (4.2)$$

The weighted pseudo-likelihood re-weights the likelihood contribution for each unit to approximate the balance of information in the target finite population, correcting for informative sampling. Estimation of the population parameters $\theta$ can proceed by using posterior samples of $\theta$. Care must be taken to ensure valid variance estimation. Since the weights are treated as fixed, population generation uncertainty is not accounted for, resulting in posterior credible intervals that do not meet frequentist nominal coverage (Gunawan et al., 2020; Léon-Novelo & Savitsky, 2019). To adjust for this, Williams & Savitsky, (2021) propose a post-processing adjustment that involves projection of the posterior samples, implemented through resampling and computation of the posterior Hessian and Jacobian, to achieve asymptotically correct coverage of credible intervals. If the sampling design includes clustering, cluster membership of observed subjects is necessary to implement the post-processing adjustment. Léon-Novelo & Savitsky (2019) also propose an alternative, fully Bayesian approach that gives efficient posterior intervals that achieve nominal coverage but requires specification of an additional Markov chain Monte Carlo (MCMC) sampler.

## 4.3 Advantages and disadvantages

The pseudo-posterior approach is advantageous in its implementation simplicity and its ability to perform parameter estimation for a large class of parametric population models, including regression and mixture models, as well as complex survey designs. MCMC samplers are easily adaptable to accommodate the weighted pseudo-likelihood, and minimal information about the complex design is required beyond the survey weights. In addition, under certain regularity conditions, the pseudo-posterior has been shown to be consistent under single-stage unequal probability sampling (Savitsky & Toth, 2016), clustered sampling (Williams & Savitsky, 2020), and complex multi-stage sampling (Williams & Savitsky, 2021).

Where the approach can run into issues is under settings with highly variable or noisy weights (Léon-Novelo & Savitsky, 2019). In those cases, the posterior variance may be reduced using strategies such as "weight smoothing", where the raw weights are replaced by their conditional expectation, obtained by regressing the weights on the outcome variables (Savitsky & Toth, 2016). Another disadvantage is that a post-processing adjustment is necessary to achieve asymptotically correct coverage of credible intervals, described and implemented in Williams & Savitsky (2021). This post-processing step is specific to each model and survey design and implementation can be particularly challenging if Stan (Carpenter et al., 2017) cannot be used for model specification. The weighted pseudo-posterior may also need to be manually derived, and analysis conclusions are dependent upon correct parametric model specification.

## 4.4 Applications and extensions

Applications and extensions of WP methods, summarized in **Table 4.1** span multiple substantive domains, including sexual health (Bastos et al., 2018; Kunihama et al., 2016, 2019), environmental health (Vedensky et al., 2022), physical and mental health (Aliverti et al., 2022; Parker et al., 2022; Parker & Holan, 2022; Trendtel & Robitzsch, 2021; Vedensky et al., 2022; Williams & Savitsky, 2021), poverty and housing (Fourrier-Nicolai & Lubrano, 2020; Gunawan et al., 2020), and employment (Savitsky et al., 2022; Savitsky & Srivastava, 2018; Savitsky & Toth, 2016; Sun et al., 2022). Methodologically, applications and extensions of the method can be broadly classified into regression models, hierarchical models, mixture models, and computational and privacy guards. The ability of WP methods to provide valid, efficient inference for a variety of parametric models holds promise for broad utility in more complex Bayesian models exploring population heterogeneity and precision health (De Vito et al., 2021; Li et al., 2021; Stephenson et al., 2022; Stephenson, Herring, et al., 2020).

**Table 4.1 Summary of applications and extensions using weighted pseudo-likelihood approach**

| | Author(s) | Data application | Survey type | Statistical method |
|---|---|---|---|---|
| **Sexual health** | Bastos et al. (2018)* | Prevalence of HIV, Hepatitis C, Hepatitis B, syphilis among transgender women in Brazil | Divas Research: Respodent-driven sampling in 12 Brazilian cities | Simple logistic regression model with weights formed from degree in respondent-driven sample |
| | Kunihama et al. (2016) | Total number of sexual partners among adolescents in the US | National Longitudinal Study of Adolescent Health: stratified sampling design | Dirichlet process mixture model with a rounded kernel method for latent continuous variables |
| | Kunihama et al. (2019)** | Trajectories of variables for adolescent sexual development in the US | National Longitudinal Study of Adolescent Health: stratified sampling design | Trajectories of associations between mixed scale longitudinal responses |
| **Envir health** | Vedensky et al. (2022) | Lead concentrations in moss in Galicia, Spain | Heavy metal biomonitoring data from Diggle et al. (2010): preferential, lattice sampling | Posterior mean predicted surfaces for geostatistical data |
| **Physical and mental health** | Aliverti & Russo (2022)* | Evolution of behaviors in compliance with COVID-19 preventive measures during lockdown in Italy | Imperial College London YouGov Covid 19 Behaviour Tracker Data Hub: repeated cross-sectional sampling | Dynamic latent-class regression model for longitudinal multivariate categorical data |
| | Parker et al. (2022)** | Health insurance rates by county in the US | American Commuity Survey: modified to informative sampling | Small area estimation of binary and count data, accounting for spatial dependence |
| | Parker & Holan (2022)** | Association between physical activity and five-year mortality in the US | National Health and Nutrition Examination Survey: stratified, multistage design | Functional regression model with binomial or multinomial outcomes |
| | Trendtel & Robitzsch (2021)** | Influence of test item position on test performance among students in New | Programme for International Student Assessment: balanced incomplete block | Probit mixed effects regression model. Balanced repeated replication for variance |
| | Williams & Savitksy (2021)** | Association between smoking and depression for adults in the US | National Survey on Drug Use and Health: stratified, multistage design | Simple logistic regression model |
| **Poverty & housing** | Fourrier-Nicolai & Lubrano (2020) | Child poverty incidence, intensity, and inequality in Germany | German Socio-Economic Panel: stratified sample with weights | Zero-inflated log-normal mixture model estimating the Three I's of Poverty (TIP) curve |
| | Gunawan et al. (2020) | Distribution of household disposable income in Australia | Household Income and Labour Dynamics in Australia: multistage sampling design | Finite gamma mixture model |
| **Employment** | Savitsky & Srivastava (2018) | Factors influencing number of employees at business establishments in California | Current Employment Statistics: informative, stratified sampling | Count regression model for large-scale data with unit missingness. Involves sample splitting |
| | Savitsky & Toth (2016)* | Factors influencing number of job hires and separations in the US | Job Openings and Labor Turnover Survey: informative sampling | Multivariate count regression model with over-dispersion |
| | Savitsky et al. (2022) | Preserving differential privacy guards for family income in the US | Current Employment Statistics: informative, stratified sampling | Generation of synthetic databases that downweight high-risk records |
| | Sun et al. (2022)** | Prevalence estimates for state-level employment and housing outcomes in the US | Household Pulse Survey: complex, multistage design | Small area estimation of binomial or multinomial outcome. Spatially correlated random effects |

*: software referenced. **: software referenced and code available

# 5. GENERATING SYNTHETIC POPULATIONS FOR COMBINING SURVEYS

## 5.1 Background and overview

Approaches that utilize synthetic population generation (SPG) to account for survey design modify the data rather than the model. Non-sampled units in the finite population of interest are treated as missing data with values that can be imputed using the observed sample data. During this imputation process, survey design elements are adjusted for, allowing the generated synthetic population to approximate the true population. Subsequent analyses can proceed without needing further survey adjustments, so standard statistical models can be applied. Note that the Bayesian aspect of this approach lies in the synthetic population generation process, but subsequent analyses of the data can use both frequentist and Bayesian methods. Synthetic populations can be used to address problems relating to disclosure risk (Little, 1993b; Raghunathan et al., 2003) and combining data from multiple surveys (Dong et al., 2014b; Raghunathan et al., 2007).

Classical multiple imputation (MI) (Rubin, 1987) techniques from the missing data literature have been extended to handle missing data (item nonresponse) within a survey context. For example, Quartagno et al. (2020), Seamen et al (2012) and Kalpourtzi et al. (2023) use multiple imputation and inverse probability weighting to incorporate sampling weights and survey design features into the imputation procedure for subjects with missing covariates or outcomes. When applied to Bayesian finite population inference, multiple imputation can be used to account for unit nonresponse and other potential differences between the observed sample data and the larger target population. This is done by generating multiple synthetic populations, which are repeatedly drawn from the posterior predictive distribution of the non-sampled units conditional on the observed data (Dong et al., 2014a; Zhou et al., 2016a). To avoid invalid inferences due to parametric imputation model misspecification (Little, 2004), a nonparametric finite population Bayesian bootstrap based on empirical likelihoods can be used (Ghosh & Meeden, 1983; Lo, 1988), with weights incorporated to account for survey design (Cohen, 1997; Dong et al., 2014a). Though not explored here, there are alternative SPG approaches that generate predictive values for the non-sampled units rather than replicate observed data as in the bootstrap-based approaches. These outcome-dependent methods assume normally distributed posterior predictive distributions and make use of Dirichlet process mixture models for robustness, with estimation primarily focused on means and totals or linear and quantile regression models (Elliott & Xia, 2021; H. J. Kim et al., 2021).

## 5.2 Description of synthetic population generation approach

Dong et al. (2014a) propose a nonparametric method for imputing the non-sampled units to generate synthetic populations that "undo" complex sampling design features, enabling subsequent analyses to proceed as if working with IID data. The approach uses a two-step procedure, where the first step resamples from the observed data to account for sampling variability, stratification, and clustering, and the second step uses a weighted sampling scheme to account for unequal selection probabilities. The method is based on the empirical likelihood of the sample data, which follows a multinomial distribution. Let $(d_1, \ldots, d_K)$ denote the set of $K$ distinct observed values of the sample outcomes $(y_1, \ldots, y_n)$, and let $\boldsymbol{\lambda} = (\lambda_1, \ldots, \lambda_K)$ denote the corresponding vector of unknown probabilities for each value in the population, with $\sum_{k=1}^{K} \lambda_k = 1$. Let $n_k$ and $u_k$ be the respective number of sampled and non-sampled units with value $d_k$, $k = 1, \ldots, K$, with $\sum_{k=1}^{K} n_k = n$ and $\sum_{k=1}^{K} u_k = N - n$. $n_k$ is known but $u_k$ must be predicted.

*Step 1: Account for sampling variability, stratification, and clustering via Bayesian bootstrap.* A Bayesian bootstrap (BB) (Rubin, 1981) is used to generate $L$ replicate samples with corresponding bootstrap replicate weights (Rao & Wu, 1988), with $L \geq 100$ for stable estimation (Dong et al., 2014a). This step captures sampling variability and reflects the uncertainty in $\boldsymbol{\lambda}$ through the varying counts of units in the original sample being selected in the replicate samples. Clustering and stratification are also handled at this stage. For cluster sampling, entire clusters are resampled rather than individual units. For stratified sampling, the resampling procedure occurs separately for each stratum. For example, for stratified cluster sampling, entire clusters are resampled, and this procedure is repeated for each stratum.

*Step 2: Undo sampling weights via weighted finite population Bayesian bootstrap.* For each of the $L$ replicate samples created in Step 1, a synthetic population is generated using the weighted finite population Bayesian bootstrap (WFPBB) (Cohen, 1997; Dong et al., 2014a). The WFPBB imputes the non-sampled units and reverses the unequal sampling design by using a weighted posterior predictive distribution given the observed sample and survey weights. To obtain the posterior predictive distribution, a weighted pseudo-empirical multinomial likelihood is formed, $p_w(n_1, \ldots, n_K | \boldsymbol{\lambda}, w_1, \ldots, w_n) \propto \prod_{k=1}^{K} \lambda_k^{w_k^*}$, where $w_k^* = \frac{n}{N-n} \sum_{i=1}^{n} (w_i - 1)^{I(y_i = d_k)}$ is the sum of the survey weights minus one for all sampled units with value $d_k$, with $\sum_{k=1}^{K} w_k^* = n$ and $I(\cdot)$ denoting the indicator function. This removes all certainty units sampled with weight one from the likelihood, as they contribute no information about non-sampled units. Then, assume a noninformative Haldane prior, $\lambda \sim Dir(0, \ldots, 0)$ (Ghosh & Meeden, 1983). The posterior predictive distribution of counts in the non-sampled data is given by

$$p(u_1, \ldots, u_K | w_1^*, \ldots, w_K^*) = \frac{\prod_{k=1}^{K} \Gamma(w_k^* + u_k)/\Gamma(w_k^*)}{\Gamma(N)/\Gamma(n)}, \qquad (5.1)$$

with $\Gamma(\cdot)$ denoting the Gamma function. Draws from this posterior predictive distribution can be easily implemented using a weighted Pólya urn sampling scheme (Cohen, 1997; Dong et al., 2014a). Available software includes the function wtpolyap in R package polyapost. Combining $n_1, \ldots, n_K$ and $u_1, \ldots, u_K$ together creates the values for the synthetic population. If the population is very large, a smaller population can be generated if it is at least 20 times as large as the sample (Dong et al., 2014a).

*Estimation and inference.* Estimation and inference for parameters proceeds by calculating estimates using each of the synthetic populations and pooling from the posterior predictive distribution. Oftentimes, this can be computationally cumbersome, and an approximation approach can be used which requires a smaller number of synthetic populations and relies on an extension of standard MI combining rules (Dong et al., 2014a; Rubin, 1987; Rubin and Raghunathan, 2003). Special care must be taken to ensure the form of the variance is correct and that the parameters of interest meet the assumptions needed for the MI combining rules to apply.

## 5.3 Advantages and disadvantages

Advantages of the SPG approach include allowing standard models to be used for analysis, having an imputation procedure that is robust, easy to implement, and not dependent on a specified outcome, and being able to handle missing data as well as combining data across multiple surveys (Dong et al., 2014a, 2014b; Zhou et al., 2016a). The approach also gives rise to unbiased point estimation and approximately unbiased variance estimation, and it requires minimal information about the complex design beyond the survey weights (Dong et al., 2014a).

Disadvantages include potential loss of efficiency compared to parametric imputation approaches, high storage burden, computational intensiveness for complicated analysis models since the model must be fit on every synthetic population, and potentially poor small sample performance if the assumption of an approximately normal repeated sampling distribution of the observed statistic is not met (Dong et al., 2014a; Zhou et al., 2016a). In addition, bias may occur if the sample inclusion mechanism depends on non-sampled outcomes, equivalent to violating the ignorable missingness assumption (Rubin, 1987).

## 5.4 Applications and extensions

Applications and extensions of SPG methods, summarized in **Table 5.1**, span multiple substantive domains, including health and injury (Zhou et al., 2016b, 2016c), healthcare coverage (Dong et al., 2014a, 2014b), poverty and income (Cocchi et al., 2022; Gunawan et al., 2020, 2021; Makela et al., 2018; Zhou et al., 2016a), and education (Goldstein et al., 2018). Methodologically, synthetic populations are used for estimation of population means, regression models, and mixture models. Though the computational burden may be prohibitive for complex models, the SPG approach holds promise for its ability to combine multiple surveys together, address missing data, and allow different analysis models to be applied without further survey adjustments post-imputation.

**Table 5.1 Summary of applications and extensions using the synthetic population generation approach**

| | Author(s) | Data application | Survey type | Statistical method |
|---|---|---|---|---|
| **Health and injury** | Zhou et al. (2016c) | Mean deceleration velocity and odds ratio of injury given varying levels of Delta-V among car crashes in the US | National Automotive Sampling System - Crashworthiness Data System: cluster, unequal probability sampling | Mean estimation and logistic regression estimation, accounting for item nonresponse |
| | Zhou et al. (2016b)** | BMI percentiles among children in the US | National Health and Nutrition Examination Survey: stratified, multi-stage sampling with oversampling | Estimation of quantiles and proportions and logistic regression model, accounting for item nonresponse |
| **Healthcare coverage** | Dong et al. (2014a) | Health insurance coverage rates overall and in demographic subpopulations in the US | National Health Interview Survey (NHIS) and Medical Expenditure Panel Survey (MEPS): stratified, multistage designs with oversampling | Population proportion estimation |
| | Dong et al. (2014b) | Health insurance coverage rates overall and in demographic subpopulations in the US | NHIS, MEPS, and Behavioral Risk Factor Surveillance System: stratified sampling, calibration | Population proportion estimation, combining multiple surveys and accounting for missing data |
| **Poverty and income** | Cocchi et al. (2022) | Mean feeding, clothing, and leisure expenses in households in Spain | Dual-frame dataset with a landline frame and a cellphone frame (R package 'Frames2') | Population mean estimation combining multiple survey frames |
| | Gunawan et al. (2020) | Distribution of household disposable income in Australia | Household, Income and Labour Dynamics in Australia: stratified, multistage panel survey with attrition | Finite gamma mixture model |
| | Gunawan et al. (2021) | Comparison of income distributions in Australia over time using stochastic dominance | Household, Income and Labour Dynamics in Australia: stratified, multistage panel survey with attrition | Infinite gamma mixture model |
| | Makela et al. (2018)* | N/A (synthetic outcome) | Fragile Families and Child Wellbeing: modified to two-stage cluster sampling with highly variable weights | Estimation of population mean for continuous or binary outcome. Account for cluster sampling by predicting unknown cluster sizes |
| | Zhou et al. (2016a) | Distribution of income and health insurance accessibility, and factors associated with BMI in the US | Behavioral Risk Factor Surveillance System: stratified sampling, calibration | Estimation of means, linear regression, and log-linear regression, accounting for item nonresponse |
| **Education** | Goldstein et al. (2018) | Factors affecting school readiness in the UK | Millennium Cohort Study: stratified, cluster, nonresponse | Multiple linear regression model, accounting for missing data |

*: software referenced. **: software referenced and code available

# 6. FUTURE DIRECTIONS FOR HEALTH DISPARITIES RESEARCH

Health disparities research hinges on the reliance of methods that are able to identify differences between subgroups within a larger population. It examines problems of inequity, where certain groups in the population are disproportionately burdened and under-resourced. Information on these at-risk groups is typically scarce, with challenges arising from reasons such as historical exclusion, accessibility issues, institutional mistrust, language barriers, and chronic disorders. This can present methodological challenges to stable and valid estimation for sparse data. Even when data is available, if the study composition differs from the population composition, care must be taken to ensure that contributions to parameter estimation reflect the composition of the true target population.

We see this consideration in MRP approaches. They are effective for utilizing data from large, national surveys to understand how basic variable estimates differ among different subgroups via small area estimation. MRP can facilitate exploration of where disparities may lie within a population by providing preliminary summaries of health outcomes in understudied populations of interest. Many of the MRP approaches applied to health disparities have centered around uses of Bayesian disease mapping in small-sized or minority groups (Malec & Müller, 2008; Kane, 2022; Zhang et al., 2021). However, further research can be done to facilitate integrating different survey datasets, examining trends over time, and accommodating other Bayesian models that examine demographic disparities in health outcomes among marginalized and hard-to-reach populations.

Mixture models and multilevel regression models have found great flexibility and utility under a Bayesian framework to identify latent profiles in diverse populations (Alexander et al., 2017; De Vito et al., 2021; Stephenson, Sotres-Alvarez et al., 2020; Stephenson & Willett, 2023; Yang & Puggioni, 2021). Such parametric models can be easily extended to account for a wide range of survey designs under the WP approach, due to its computational convenience. These extensions will permit more complex questions about population health dynamics to be answered while maintaining generalizability. They hold much potential in population health disparity applications, given their ability to provide valid and efficient inference for parametric models for a wide variety of data types.

SPG finds its strength in in its ability to accommodate missing data and information pooled from multiple data sources. These issues are commonly encountered when working with understudied populations and survey studies with significant nonresponse. The implementation of SPG techniques allows for standard analysis models to be applied to modified (augmented) data with readily available computational tools and resources. Similarly, the combination of multiple surveys may be particularly useful for research on underrepresented or understudied groups. In many cases, there is a lack of data, and no single large, representative survey exists with enough power to adequately answer questions of interest for target demographic; instead, various smaller surveys capturing different relevant aspects must be combined for analysis.

A major limitation of survey studies aimed at examining population health disparities is accounting for nonresponse. Response rates have continued to decline over time (Czajka & Beyler, 2016; Stedman et al., 2019). Further, response rate patterns can differ for certain subgroups depending on study scope and aims (Sjöström & Holst, 2009). These challenges, coupled with an increasing availability of alternative data sources (e.g., web-based surveys, databases, and community-based convenience samples), have led to research gaining traction in the area of survey inference for non-probability samples (S. Chen et al., 2022; Y. Chen et al., 2020; J. K. Kim et al., 2021; Rafei et al., 2022). Non-probability samples are useful for accessing rare and hard-to-reach populations and can be very cost-effective, though extra care must be taken when generalizing results. In non-probability samples, there are no traditional survey weights available because the selection mechanism is no longer known by design. Instead, differences between the sample and population must be accounted for by either: 1) estimating pseudo-survey weights by modeling the selection mechanism and then forming an inverse probability weighted estimator, as in quasi-randomization approaches; 2) conditioning out variables that influence selection in an outcome model for prediction of non-sampled units, as in superpopulation modeling methods; or 3) a combination of both, as in doubly robust approaches. Elliott and Valliant detail current inferential methods for non-probability samples (Elliott & Valliant, 2017; Valliant, 2020). Most methods in this area focus on estimation of the population mean and derive from the causal inference literature, repurposing tools used to control for confounding to now address variables influencing selection (Mercer et al., 2017). These methods also require a reference probability sample or auxiliary population data to properly calibrate the non-probability sample and avoid bias (Thompson, 2019). While many methods still fall within the frequentist paradigm, there are some recent advances that have incorporated Bayesian modeling (Rafei et al., 2020, 2022; Tan et al., 2019). The Bayesian approaches discussed in this review can also be extended to accommodate non-probability samples under the quasi-randomization approach for WP and a model-based superpopulation approach for MRP and SPG.

## Conclusion

In this review, we discussed Bayesian methods for analyzing survey data to motivate opportunities to extend these approaches to potential population health disparities applications. Surveys have provided an effective way to bridge the gap of informational access to understudied populations and relatively low-cost and non-invasive data collection procedures. They are often incorporated into routine public health surveillance and provide consistent, updated information on populations of interest. In accommodating large population heterogeneity, Bayesian methods offer advantages for flexibly adapting to sparse and complex survey data within a coherent inferential framework.

The three main Bayesian survey approaches explored were: 1) multilevel regression and post-stratification (MRP), 2) weighted pseudo-likelihood (WP), and 3) synthetic population generation (SPG). A summary of these approaches is provided in **Table 6.1**. MRP combines a Bayesian multilevel regression model with post-stratification to allow for stable subgroup estimation while accounting for survey design. MRP is an efficient method for estimating means and totals of geographic and demographic subgroup characteristics but does not handle more complex parametric analyses. WP accounts for survey design in parametric models by forming a weighted pseudo-likelihood with the survey weights and conducting estimation using a pseudo-posterior. It is computationally convenient for adapting a wide array of parametric models to account for complex survey designs. SPG adjusts for complex survey design by creating synthetic populations from multiply imputed non-sampled units and running parallel analysis on each bootstrapped population, pooling the results to obtain estimates. It allows multiple analyses to be applied to the same set of synthetic populations and is useful for handling missing data and combining data sources. Computationally, these approaches vary considerably. Publicly available software and tutorials are available for MRP methods. WP methods require derivation modification, but examples and code have been made available for reference (Williams & Savitsky, 2021). SPG methods have been found to be somewhat limited in available software.

**Table 6.1 Summary of advantages and disadvantages for the main approaches**

| Approach | Advantages | Disadvantages |
|---|---|---|
| Multilevel regression and post-stratification (MRP) | + Accurate and efficient for small area estimation<br>+ Stable in the presence of sparse data<br>+ Quick and easy to implement using Stan<br>+ Only requires information on the weights and cluster membership | - Requires additional pooling if sparse sample cells<br>- Model specification can be challenging<br>- Limited to univariate descriptive inference of means and totals<br>- Assumes ignorable missingness |
| Weighted pseudo-likelihood (WP) | + Consistency under complex sampling<br>+ Computationally convenient and relatively simple to implement for models using Stan<br>+ Applicable to many parametric models<br>+ Only requires information on the weights and cluster membership | - Sensitive to highly variable weights<br>- Weighted pseudo-posterior may require derivation<br>- Requires additional adjustment for valid variance estimation, which may be difficult to implement |
| Data augmentation (DA) | + Standard analysis models can be used<br>+ Imputation procedure is easy to implement and not dependent on outcome<br>+ Handles missing data and combining multiple data sources<br>+ Unbiased point and variance estimation<br>+ Only requires information on the weights and cluster membership | - Less efficient than other approaches<br>- Requires storage of many synthetic populations<br>- Computationally intensive for complicated models<br>- Small sample performance may be poor<br>- Assumes ignorable missingness |

The integration of Bayesian statistics and survey methodology has been slow and often met with resistance. Much work in this area remains to be done. There is a great opportunity to improve health disparities research by building upon the current tools available and developing extensions to these methods. MRP can improve population-based estimates of health effects derived for smaller-sized or underrepresented subgroups. WP can improve inference on understudied populations through its implementation in flexible and parametric models. SPG can amplify underrepresented populations by pooling together multiple studies in an effort to increase power and overcome missing data challenges. Further, integration of these methods with non-probability samples will allow these tools to remain relevant as modern data sources become more available and increase our coverage of hard-to-reach populations and geographic areas. Publicly available and easy-to-use software is relatively scarce, but increased awareness of this gap should motivate researchers to develop computational tools for a broad audience of researchers engaged in achieving population health equity.

**Funding Information**

This research was supported in part by the National Institute of Allergy and Infectious Diseases (NIAID: T32 AI007358) and the National Heart, Lung, and Blood Institute (NHLBI: R25 HL105400 awarded to Victor G. Davila-Roman and DC Rao).

**Acknowledgments**

The authors would like to thank the researchers who have mobilized the field of Bayesian survey statistics to this point and the editors for providing the opportunity to share these exciting research developments and future directions.